\documentclass[aps,pre,twocolumn,superscriptaddress,nofootinbib,longbibliography]{revtex4-2}

\usepackage{amsmath,amssymb,amsfonts,bm}
\usepackage{mathtools}
\usepackage{graphicx}
\usepackage{microtype}
\usepackage{hyperref}
\hypersetup{colorlinks=true,linkcolor=blue,citecolor=blue,urlcolor=blue}

\newcommand{\ep}{\dot e_{\mathrm p}}
\newcommand{\hdr}{\dot h_{\mathrm d}}

\newcommand{\st}{\mathrm{st}}
\newcommand{\ssv}{\mathrm{ss}}
\newcommand{\avg}[1]{\left\langle #1\right\rangle}
\newcommand{\dd}{\mathrm d}
\newcommand{\e}{\mathrm e}

\begin{document}

\title{Force--Torque Reciprocity and the Inference of Concealed Dissipation in a Geared Brownian Machine}

\author{Mesfin Asfaw Taye}
\email{tayem@wlac.edu}
\affiliation{West Los Angeles College, Science Division, 9000 Overland Ave., Culver City, California 90230, USA}

\date{\today}

\begin{abstract}
We show that in a reciprocal Brownian motor the entropy production hidden behind a mechanically stalled coordinate can be reconstructed exactly from measurements of that coordinate alone.  We introduce a minimal, analytically solvable Langevin motor in which an observed translational coordinate is coupled reciprocally to a hidden internal rotor: a single periodic potential $V(x-\ell\theta)$ generates both the force on the observed coordinate and the reaction torque on the hidden one, so that $\tau_{\rm int}=-\ell F_x$ holds identically.  Force--torque reciprocity together with translational symmetry produces a local current identity that closes the hidden thermodynamic bookkeeping.  From it we prove that the Harada--Sasa heat measured through the observed coordinate equals the positive current-square dissipation of that coordinate, with the information-flow correction vanishing identically.  This yields, at mechanical stall, an exact reconstruction of the full entropy production from the fluctuation--response violation $\Phi_x^{\st}$ of the observed coordinate when the reciprocal mobility factor $K=1+\ell^2\mu_\theta/\mu_x$ is calibrated,
\begin{equation*}
\ep^{\st}=K\Phi_x^{\st}+\frac{TK^2(\Phi_x^{\st})^2}{\mu_x(K-1)f_{\st}^2},
\end{equation*}
and an optimized model-class bound, $\ep^{\st}\geq\Phi_x^{\st}\big(\sqrt{1+a}+\sqrt a\big)^2$ with $a=T\Phi_x^{\st}/\mu_xf_{\st}^2$, that is strictly stronger than the direct visible-dissipation bound and is saturated by an explicit member of the reciprocal class.  All results are confirmed both by direct quadrature of the stationary Fokker--Planck solution and by Langevin simulation of the original two-variable dynamics.  The construction provides an analytically solvable benchmark for entropy-production inference from a single observed coordinate, distinct from earlier time-scale-separation recoveries in that it requires no separation of time scales.
\end{abstract}

\maketitle

\section{Introduction}
\label{sec:intro}

Nonequilibrium thermodynamics and statistical mechanics remain challenging because systems far from equilibrium lack the universal description that the Boltzmann--Gibbs distribution provides at equilibrium.  Central open problems include quantifying microscopic entropy production, predicting steady-state distributions, and accounting for the role of fluctuations~\cite{GeQian2010,TomeOliveira2012,Schnakenberg1976,TomeOliveira2010,ZiaSchmittmann2007}.  Over the past two decades a robust framework has nonetheless emerged.  The dependence of entropy production on model parameters has been analyzed through the master equation for discrete systems~\cite{GeQian2010,Schnakenberg1976,Tome2006,SzaboTome2010} and through the Fokker--Planck equation for continuous systems~\cite{Seifert2005,Taye2016,Taye2020,TomeOliveira2015}.  Various thermodynamic relations have been derived using stochastic thermodynamics~\cite{Seifert2005,Seifert2012}, time-reversal operations~\cite{Ge2014,LeeKwonPark2013}, and the fluctuation theorem~\cite{LebowitzSpohn1999,SpinneyFord2012,Celani2012}, while the thermodynamic uncertainty relation has more recently provided a powerful route to estimating entropy production from time-series data~\cite{Manikandan2020,SkinnerDunkel2021,Otsubo2020,KoyukSeifert2019}.  Entropy production has thus become a central observable through which the irreversibility of a driven system is read off from its dynamics.

The entropy production is built from the full set of microscopic probability currents and their conjugate forces.  Experiment, however, rarely has access to the full current field.  One typically measures a single coordinate---the displacement of a colloid, the position of a bead in an optical trap, the step of a molecular motor along its track---while many chemical, conformational, or otherwise internal degrees of freedom remain hidden.  The measured quantity is then a projection of the underlying dynamics.

It is by now well understood that a vanishing measured current does not imply reversibility.  At mechanical stall, where the mean displacement and the output power vanish, a system driven by chemical free energy and opposed by an external load can nonetheless remain dissipative, because the unmeasured currents need not vanish.  This qualitative fact has been established in several settings: hidden cycle currents in discrete Markov networks~\cite{Esposito2010,Maes2003,Taye2026}, and inference of broken detailed balance and entropy production from partial trajectories using waiting-time statistics, current fluctuations, and fluctuation--response violation~\cite{HaradaSasa2005,HaradaSasa2006,Bisker2017,Martinez2019}.  Closest to the present work, Wang, Kawaguchi, Sasa, and Tang~\cite{Wang2016} showed, in a solvable time-scale-separated model, that hidden entropy production can be recovered through the fluctuation--response spectrum of a slow observable, with the spectrum fully accounting for the dissipation generated by fast switching.  More generally, for coupled (bipartite) subsystems the entropy balance of a single subsystem is modified by an information-flow term relative to that subsystem's heat dissipation~\cite{HorowitzEsposito2014}.  These studies establish that a stalled coordinate can hide dissipation and that fluctuation--response violation can flag its presence.  What has remained open is the quantitative question: under what structural conditions do measurements of the stalled coordinate alone determine the full hidden dissipation, rather than merely signal that it is nonzero?  This is the question we resolve.

The present contribution is quantitative, and rests on a structural assumption that the prior partial-observation literature does not exploit.  In a generic hidden-variable model the hidden coordinate is an externally prescribed clock that drives the observed one without back-reaction, so the energy balance is not closed and the hidden dissipation cannot be recovered from observed data alone.  We instead require the hidden coordinate to feel the mechanical reaction of the very interaction that drives the observed coordinate.  Concretely, an observed translational coordinate $x$ is coupled to a hidden internal phase $\theta$ through a single periodic potential $V(x-\ell\theta)$, so that the force on $x$ and the reaction torque on $\theta$ are tied identically by $\tau_{\rm int}=-\ell F_x$.  We call this property \emph{force--torque reciprocity}; it is a mechanical reciprocity arising from a shared potential, not an Onsager relation restricted to linear response.  Because the dynamics depends only on the relative coordinate
\begin{equation}
q=x-\ell\theta,
\label{eq:q_intro}
\end{equation}
the steady state reduces exactly to diffusion in a tilted periodic potential of the type long used to model Brownian heat engines and ratchets~\cite{Taye2022,Taye2024,Reimann2002,Hanggi2009}, and all steady currents and the entropy production follow in closed form.  The combination $q$ is the relative mechanical phase, or slip, between the external displacement and the internal cycle: when $q$ remains localized the two coordinates are tightly geared, and when $q$ circulates the rotor slips relative to the track.  This makes the stall mechanism transparent---at mechanical stall $x$ has no mean motion, but a rotating hidden phase produces a nonzero slip velocity $\avg{\dot q}=-\ell\avg{\dot\theta}$ and continues to dissipate.

Our answer rests on force--torque reciprocity and translational symmetry, which together yield a potential-independent local current identity at mechanical stall that closes the hidden thermodynamic bookkeeping.  Crucially, this requires no separation of time scales between the observed and hidden coordinates---in contrast to the fluctuation--response recoveries available only in time-scale-separated models~\cite{Wang2016}.  The identity shows that the Harada--Sasa heat measured through the observed coordinate equals its positive current-square dissipation, while the information-flow correction normally present in coupled subsystems~\cite{HorowitzEsposito2014} vanishes identically.  Consequently, the visible fluctuation--response violation determines the hidden dissipation: the full entropy production is reconstructed exactly once $K=1+\ell^2\mu_\theta/\mu_x$ is calibrated, and otherwise obeys an optimized model-class bound sharper than the direct visible-dissipation bound.  We also identify a reciprocal model that saturates this bound and verify all relations by stationary Fokker--Planck quadrature and direct Langevin simulation.  This potential-independent reciprocal stall identity, and the exact single-coordinate reconstruction it permits without time-scale separation, are the central contributions of this work.

The principal result is an inference theorem.  Let $f_{\st}$ denote the mechanical-stall load and let
\begin{equation}
\Phi_x^{\st}=\frac{1}{\mu_xT}\int_{-\infty}^{\infty}\frac{\dd\omega}{2\pi}
\left[C_{vv}(\omega)-2T\,\mathrm{Re}\,R_v(\omega)\right]
\label{eq:intro_phi}
\end{equation}
be the Harada--Sasa violation of the observed velocity $v=\dot x$ at stall.  If the reciprocal mobility factor
\begin{equation}
K=1+\frac{\ell^2\mu_\theta}{\mu_x}
\label{eq:intro_K}
\end{equation}
is independently calibrated, the full stall entropy production is
\begin{equation}
\ep^{\st}=K\Phi_x^{\st}
+\frac{TK^2(\Phi_x^{\st})^2}{\mu_x(K-1)f_{\st}^2}
\label{eq:intro_recon}
\end{equation}
for $\ell f_{\st}\neq0$.  When the hidden mobility is unknown, minimization over all $K>1$ gives the attainable reciprocal bound
\begin{equation}
\ep^{\st}\geq \Phi_x^{\st}\left(\sqrt{1+a}+\sqrt a\right)^2,
\qquad
 a=\frac{T\Phi_x^{\st}}{\mu_x f_{\st}^2}.
\label{eq:intro_bound}
\end{equation}
The bound is stricter than $\ep^{\st}\geq\Phi_x^{\st}$ because reciprocal motion in the observed channel necessarily produces motion and dissipation in the hidden channel as well.

The article is organized as follows.  Section~\ref{sec:modelthermo} defines the reciprocal motor and derives its thermodynamic balance.  Section~\ref{sec:exactstall} solves the steady state, identifies mechanical stall, and distinguishes it from detailed balance.  Section~\ref{sec:inference} derives the observable dissipation, the exact reconstruction, and the optimized bound.  Section~\ref{sec:verification} develops useful analytical limits and verifies the results numerically.  Section~\ref{sec:energetics} works out the full energetics, power conversion, and operating regimes of the motor, including the complete energetic reconstruction at stall.  Section~\ref{sec:discussion} discusses the physical content, scope, and experimental interpretation.  Detailed algebra that would interrupt the main argument is retained in the Appendices.

\section{Reciprocal motor and thermodynamic structure}
\label{sec:modelthermo}

\paragraph*{Mechanical construction and meaning of the variables.}
The observed coordinate $x$ is periodic with period $L$, and the hidden coordinate $\theta$ is an angle of period $2\pi$.  We use the minimal one-to-one gearing length
\begin{equation}
\ell=\frac{L}{2\pi},
\label{eq:ell}
\end{equation}
so that $q=x-\ell\theta$ is periodic with period $L$.  The hidden angle may represent a coarse-grained chemical cycle, a conformational phase, or an internal rotor.  The quantity $\ell\theta$ is the displacement that the internal cycle would generate under perfect gearing, while $q$ measures the mismatch between this preferred displacement and the observed position.

The interaction energy is assumed to depend only on the mismatch,
\begin{equation}
U(x,\theta)=V(q)=V(x-\ell\theta),
\qquad V(q+L)=V(q).
\label{eq:U}
\end{equation}
The mechanical force and the internal reaction torque generated by this one energy are
\begin{equation}
F_x=-\partial_xU=-V'(q),
\qquad
\tau_{\rm int}=-\partial_\theta U=\ell V'(q),
\label{eq:force_torque}
\end{equation}
and therefore
\begin{equation}
\tau_{\rm int}=-\ell F_x.
\label{eq:reciprocity}
\end{equation}
Equation~\eqref{eq:reciprocity} is the central structural assumption.  A force exerted by the rotor on the translational coordinate is accompanied by the opposite reaction torque on the rotor, expressed in consistent length units.  In contrast, a one-way hidden-clock model would prescribe a force $F(\theta)$ in the $x$ equation while leaving the $\theta$ dynamics unaffected; such a model is useful phenomenologically but does not close the energetic exchange between the two coordinates.

\paragraph*{Overdamped stochastic dynamics.}
A constant active torque $\tau_{\rm a}$ drives the hidden rotor, and a load $f$ opposes positive motion in $x$.  With Boltzmann's constant set to unity, the overdamped Langevin equations are
\begin{align}
\dot x &= \mu_x\left[-V'(q)-f\right]+\sqrt{2\mu_xT}\,\xi_x(t),
\label{eq:langevin_x}\\
\dot\theta &= \mu_\theta\left[\tau_{\rm a}+\ell V'(q)\right]
+\sqrt{2\mu_\theta T}\,\xi_\theta(t).
\label{eq:langevin_th}
\end{align}
The independent noises satisfy
\begin{equation}
\avg{\xi_i(t)}=0,
\qquad
\avg{\xi_i(t)\xi_j(t')}=\delta_{ij}\delta(t-t').
\label{eq:noise}
\end{equation}
The mobilities $\mu_x$ and $\mu_\theta$ convert force and torque into velocity.  Both variables contact the same heat bath at temperature $T$.  Since the noise amplitudes are constant, It\^o and Stratonovich interpretations give the same state dynamics.  Products used to define stochastic heat are understood in the Stratonovich convention, as required by ordinary calculus in stochastic energetics~\cite{Sekimoto2010}.

The load and the active torque play different thermodynamic roles.  The active torque injects power $\tau_{\rm a}J_\theta$ through the hidden cycle, whereas the load extracts power $fJ_x$ when $J_x>0$.  The periodic potential is conservative and only transfers energy between the two coordinates.  The motor branch considered below has $\tau_{\rm a}>0$, $f\geq0$, and a hidden current in the direction of the active torque.

\paragraph*{Fokker--Planck currents.}
Let $p(x,\theta,t)$ be the probability density on the torus $[0,L)\times[0,2\pi)$.  The Langevin equations imply
\begin{equation}
\partial_t p=-\partial_xj_x-\partial_\theta j_\theta,
\label{eq:fpe}
\end{equation}
with currents
\begin{align}
 j_x&=\mu_x[-V'(q)-f]p-\mu_xT\,\partial_xp,
\label{eq:jx}\\
 j_\theta&=\mu_\theta[\tau_{\rm a}+\ell V'(q)]p
-\mu_\theta T\,\partial_\theta p.
\label{eq:jth}
\end{align}
The mean translational and angular velocities are
\begin{equation}
J_x=\int_0^L\dd x\int_0^{2\pi}\dd\theta\,j_x,
\qquad
J_\theta=\int_0^L\dd x\int_0^{2\pi}\dd\theta\,j_\theta.
\label{eq:mean_currents}
\end{equation}
The notation distinguishes the local probability currents $j_i(x,\theta)$ from their spatial integrals $J_i$.  This distinction is essential: mechanical stall constrains $J_x$, while reversibility constrains both local currents pointwise.

\paragraph*{Entropy balance.}
The Gibbs--Shannon entropy is
\begin{equation}
S(t)=-\int\dd x\,\dd\theta\,p\ln p.
\label{eq:entropy}
\end{equation}
Using Eq.~\eqref{eq:fpe}, periodic boundary conditions, and normalization gives
\begin{equation}
\dot S=-\int\dd x\,\dd\theta
\left(j_x\partial_x\ln p+j_\theta\partial_\theta\ln p\right).
\label{eq:Sdot_first}
\end{equation}
The current definitions can be rearranged as
\begin{align}
\partial_x\ln p&=\frac{-V'(q)-f}{T}-\frac{j_x}{\mu_xTp},\\
\partial_\theta\ln p&=\frac{\tau_{\rm a}+\ell V'(q)}{T}
-\frac{j_\theta}{\mu_\theta Tp}.
\end{align}
Substitution into Eq.~\eqref{eq:Sdot_first} separates the entropy change into production and extraction,
\begin{equation}
\dot S=\ep-\hdr,
\label{eq:entropy_balance}
\end{equation}
where
\begin{equation}
\ep=\int\dd x\,\dd\theta
\left[\frac{j_x^2}{\mu_xTp}+\frac{j_\theta^2}{\mu_\theta Tp}\right]\geq0
\label{eq:ep}
\end{equation}
and
\begin{equation}
\hdr=\frac{1}{T}\int\dd x\,\dd\theta
\left\{j_x[-V'(q)-f]+j_\theta[\tau_{\rm a}+\ell V'(q)]\right\}.
\label{eq:hd}
\end{equation}
Every integrand in Eq.~\eqref{eq:ep} is nonnegative.  Consequently, $\ep=0$ if and only if $j_x=j_\theta=0$ everywhere.  At a steady state $\dot S=0$, so entropy production equals entropy extraction, $\ep^{\ssv}=\hdr^{\ssv}$.  This equality is a balance law, not an equilibrium condition: both rates may be positive in a nonequilibrium steady state.

\paragraph*{Power balance.}
The mean interaction energy changes according to
\begin{align}
\frac{\dd}{\dd t}\avg{U}
&=\int\dd x\,\dd\theta\left(\partial_xU\,j_x+\partial_\theta U\,j_\theta\right)\nonumber\\
&=\int\dd x\,\dd\theta\,V'(q)(j_x-\ell j_\theta).
\label{eq:Udot}
\end{align}
Using Eq.~\eqref{eq:force_torque} in Eq.~\eqref{eq:hd} yields the exact first-law form
\begin{equation}
T\hdr=-\frac{\dd}{\dd t}\avg U+\tau_{\rm a}J_\theta-fJ_x.
\label{eq:firstlaw}
\end{equation}
At steady state the bounded periodic interaction stores no net energy, and Eqs.~\eqref{eq:entropy_balance} and \eqref{eq:firstlaw} give
\begin{equation}
T\ep^{\ssv}=\tau_{\rm a}J_\theta-fJ_x.
\label{eq:power}
\end{equation}
Thus $T\ep$ is the dissipated power, $\tau_{\rm a}J_\theta$ is active input power, and $fJ_x$ is mechanical output power.  This identity is exact because the force and torque originate from the same interaction energy.

\section{Exact steady state, mechanical stall, and reversibility}
\label{sec:exactstall}

\paragraph*{Symmetry and reduction to the slip coordinate.}
The generator of Eqs.~\eqref{eq:langevin_x}--\eqref{eq:langevin_th} is invariant under the simultaneous shift
\begin{equation}
(x,\theta)\mapsto(x+\ell\alpha,\theta+\alpha),
\label{eq:symmetry}
\end{equation}
which leaves $q=x-\ell\theta$ unchanged.  Finite temperature makes the diffusion on the compact torus irreducible, so the stationary density is unique.  Applying the symmetry to that unique density cannot generate a different stationary state; hence the density is constant along the symmetry direction and depends only on $q$:
\begin{equation}
p_{\ssv}(x,\theta)=\frac{1}{2\pi}\rho(q),
\qquad
\int_0^L\rho(q)\,\dd q=1.
\label{eq:factorization}
\end{equation}
The $x$ and $\theta$ marginals are therefore uniform, although the relative coordinate can have a highly nonuniform nonequilibrium density.

Subtracting $\ell\dot\theta$ from $\dot x$ gives an autonomous Langevin equation for the slip,
\begin{equation}
\dot q=-M[V'(q)+\Lambda]+\sqrt{2MT}\,\eta_q(t),
\label{eq:q_langevin}
\end{equation}
where
\begin{equation}
M=\mu_x+\ell^2\mu_\theta,
\qquad
\Lambda=\frac{\mu_xf+\ell\mu_\theta\tau_{\rm a}}{M},
\label{eq:M_Lambda}
\end{equation}
and
\begin{equation}
\eta_q=\frac{\sqrt{\mu_x}\,\xi_x-\ell\sqrt{\mu_\theta}\,\xi_\theta}{\sqrt M}
\end{equation}
is a normalized white noise.  Both positive load and positive active torque decrease $q$: the load drives $x$ backward, while the torque drives $\theta$ forward.  They therefore add in the effective tilt $\Lambda$.  The motor action arises because the periodic coupling converts part of the driven rotor motion into positive translation of $x$.

The reduced stationary current is constant,
\begin{equation}
J_q=-M[V'(q)+\Lambda]\rho(q)-MT\rho'(q).
\label{eq:Jq_local}
\end{equation}
With
\begin{equation}
\psi(q)=\frac{V(q)+\Lambda q}{T},
\end{equation}
the first-order equation can be integrated explicitly.  Periodicity gives
\begin{equation}
\rho(q)=\frac{J_q}{MT[1-\e^{\Lambda L/T}]}
\e^{-\psi(q)}\int_q^{q+L}\e^{\psi(y)}\,\dd y,
\label{eq:rho_exact}
\end{equation}
and normalization gives the exact tilted-periodic current~\cite{Risken1989,Reimann2002}
\begin{equation}
J_q(\Lambda)=
\frac{MT[1-\e^{\Lambda L/T}]}
{\displaystyle\int_0^L\dd q\,\e^{-\psi(q)}
\int_q^{q+L}\dd y\,\e^{\psi(y)}}.
\label{eq:Jq_exact}
\end{equation}
The mean slip velocity is
\begin{equation}
v_q=LJ_q=J_x-\ell J_\theta.
\label{eq:vq_def}
\end{equation}
When $\Lambda=0$, Eq.~\eqref{eq:Jq_exact} is understood by continuity: $J_q=0$ and $\rho\propto\exp[-V(q)/T]$.

\paragraph*{Reconstruction of the two mean currents.}
Averaging Eqs.~\eqref{eq:jx} and \eqref{eq:jth} with Eq.~\eqref{eq:factorization} gives
\begin{equation}
J_x=\mu_x[-\avg{V'}-f],
\qquad
J_\theta=\mu_\theta[\tau_{\rm a}+\ell\avg{V'}],
\label{eq:mean_force_currents}
\end{equation}
where $\avg{V'}=\int_0^LV'(q)\rho(q)\,\dd q$.  Eliminating $\avg{V'}$ in favor of $v_q=J_x-\ell J_\theta$ gives
\begin{align}
J_x&=\frac{\mu_x}{M}\left[v_q+\ell\mu_\theta(\tau_{\rm a}-\ell f)\right],
\label{eq:Jx_vq}\\
J_\theta&=\frac{\mu_\theta}{M}\left[\mu_x(\tau_{\rm a}-\ell f)-\ell v_q\right].
\label{eq:Jth_vq}
\end{align}
Equations~\eqref{eq:Jq_exact}, \eqref{eq:Jx_vq}, and \eqref{eq:Jth_vq} solve the complete steady transport problem for an arbitrary periodic coupling $V$.

\paragraph*{Mechanical stall and hidden slip.}
Mechanical stall is defined operationally by
\begin{equation}
J_x(f_{\st})=0.
\label{eq:stall_def}
\end{equation}
Using Eq.~\eqref{eq:Jx_vq}, the exact stall load is the root of
\begin{equation}
v_q(\Lambda_{\st})+\ell\mu_\theta(\tau_{\rm a}-\ell f_{\st})=0,
\qquad
\Lambda_{\st}=\frac{\mu_xf_{\st}+\ell\mu_\theta\tau_{\rm a}}{M}.
\label{eq:stall_condition}
\end{equation}
This is an implicit but closed analytical condition because $v_q$ is given by the quadrature \eqref{eq:Jq_exact}.  At stall,
\begin{equation}
J_\theta^{\st}=\mu_\theta(\tau_{\rm a}-\ell f_{\st}),
\qquad
v_q^{\st}=-\ell J_\theta^{\st}.
\label{eq:hidden_stall_current}
\end{equation}
The second relation gives the physical meaning of hidden dissipation: if the observed position is fixed on average while the internal phase advances, all internal motion appears as backward circulation of the relative slip.

The output power vanishes at stall, but the input power need not.  Equation~\eqref{eq:power} becomes
\begin{equation}
T\ep^{\st}=\tau_{\rm a}J_\theta^{\st}
=\tau_{\rm a}\mu_\theta(\tau_{\rm a}-\ell f_{\st}).
\label{eq:stall_ep}
\end{equation}
For $\tau_{\rm a}>0$, the second law requires $J_\theta^{\st}\geq0$.  A genuinely nonequilibrium stalled motor has $J_\theta^{\st}>0$ and therefore
\begin{equation}
0\leq f_{\st}<\frac{\tau_{\rm a}}{\ell}.
\label{eq:stall_range}
\end{equation}
The upper equality cannot occur for nonzero driving, as shown next.

\paragraph*{Detailed balance and the topology of the torus.}
Reversibility requires $j_x=j_\theta=0$ pointwise, not merely $J_x=0$.  This condition can be tested by integrating the zero-current equations around the two independent cycles of the torus.  If $j_x=0$, then
\begin{equation}
\partial_x\ln p_{\ssv}=\frac{-V'(q)-f}{T}.
\end{equation}
At fixed $\theta$, integration over $x\to x+L$ gives $0=-fL/T$, because $V$ is periodic; hence $f=0$.  Similarly, $j_\theta=0$ gives
\begin{equation}
\partial_\theta\ln p_{\ssv}=\frac{\tau_{\rm a}+\ell V'(q)}{T}.
\end{equation}
At fixed $x$, integration over one $2\pi$ rotation gives $0=2\pi\tau_{\rm a}/T$, because the integral of the periodic torque is zero; hence $\tau_{\rm a}=0$.  Therefore
\begin{equation}
\text{detailed balance holds only at }f=0,\quad\tau_{\rm a}=0.
\label{eq:db_point}
\end{equation}
At this point $\Lambda=0$, $J_q=J_x=J_\theta=0$, and $\rho\propto\exp[-V/T]$.  A nonzero load $f=\tau_{\rm a}/\ell$ may cancel the mean mismatch in Eq.~\eqref{eq:hidden_stall_current}, but it cannot cancel the two independent nonconservative affinities around the torus.  Thus it is not a reversible operating point.

The distinction can also be written geometrically.  With $\bm j_{\ssv}=(j_x,j_\theta)$ and $D=\operatorname{diag}(\mu_xT,\mu_\theta T)$,
\begin{equation}
\ep=\int\dd x\,\dd\theta\,
\frac{\bm j_{\ssv}^{\mathsf T}D^{-1}\bm j_{\ssv}}{p_{\ssv}}.
\label{eq:projection_form}
\end{equation}
The condition $\ep=0$ requires the entire vector field $\bm j_{\ssv}$ to vanish.  Mechanical stall imposes only one scalar projection, $J_x=\int j_x=0$.  A signed projection can cancel, whereas the positive quadratic form in Eq.~\eqref{eq:projection_form} cannot.

\paragraph*{Weak coupling and the origin of a nonzero stall load.}
For a sinusoidal coupling $V(q)=V_0\cos(kq)$, $k=2\pi/L$, the exact current has the small-amplitude expansion
\begin{equation}
v_q(\Lambda)=-M\Lambda\left[1-\frac{k^2V_0^2}{2[(Tk)^2+\Lambda^2]}\right]+O(V_0^4).
\label{eq:vq_weak}
\end{equation}
Since $f_{\st}=O(V_0^2)$, substitution into Eq.~\eqref{eq:stall_condition} gives
\begin{equation}
f_{\st}=
\frac{\ell\mu_\theta\tau_{\rm a}}{M}
\frac{k^2V_0^2}{2\left[(Tk)^2+(\ell\mu_\theta\tau_{\rm a}/M)^2\right]}
+O(V_0^4).
\label{eq:fst_weak}
\end{equation}
The free case $V_0=0$ has $f_{\st}=0$: an uncoupled rotor cannot perform translational work.  The coupling generates a positive stall force at second order because the first-order modulation averages to zero over a period.  This limit provides a simple physical check on the exact formulas.

\section{Observable dissipation and exact stall-point inference}
\label{sec:inference}

\paragraph*{The local reciprocal-current identity.}
The exact inference result begins with local mean velocities,
\begin{equation}
\nu_x(q)=\frac{j_x}{p},
\qquad
\nu_\theta(q)=\frac{j_\theta}{p}.
\label{eq:local_vel_def}
\end{equation}
Using $p_{\ssv}=\rho(q)/(2\pi)$, so that $\partial_x\ln p=\partial_q\ln\rho$ and $\partial_\theta\ln p=-\ell\partial_q\ln\rho$, Eqs.~\eqref{eq:jx} and \eqref{eq:jth} give
\begin{align}
\nu_x&=\mu_x[-V'(q)-f]-\mu_xT\,\partial_q\ln\rho,
\label{eq:nux_raw}\\
\nu_\theta&=\mu_\theta[\tau_{\rm a}+\ell V'(q)]
+\ell\mu_\theta T\,\partial_q\ln\rho.
\label{eq:nuth_raw}
\end{align}
The potential force and the density-gradient term cancel when the two equations are combined.  The result is the pointwise identity
\begin{equation}
\nu_\theta(q)=\mu_\theta(\tau_{\rm a}-\ell f)
-\frac{\ell\mu_\theta}{\mu_x}\nu_x(q).
\label{eq:local_identity}
\end{equation}
This relation holds for every $q$, every periodic potential, and every driving strength.  It is stronger than a relation between mean currents: once the local motion in $x$ is known, reciprocity fixes the local hidden motion up to one constant offset.

It is useful to define
\begin{equation}
K=1+\frac{\ell^2\mu_\theta}{\mu_x}=\frac{M}{\mu_x}>1.
\label{eq:K}
\end{equation}
The factor $K$ measures the hidden mobility after it is reflected into translational units by the gearing length $\ell$.  A larger $K$ means that a given local translational fluctuation is accompanied by a larger hidden response.  Since the local slip velocity is $\nu_q=\nu_x-\ell\nu_\theta$, Eq.~\eqref{eq:local_identity} also implies
\begin{equation}
\rho(q)\left[K\nu_x(q)-\ell\mu_\theta(\tau_{\rm a}-\ell f)\right]=J_q.
\label{eq:nux_closure}
\end{equation}
This closure is the mathematical reason the hidden dissipation becomes inferable.

\paragraph*{Heat in the observed channel and the correlation-flow correction.}
Let $\dot Q_x$ denote heat dissipated into the bath through the observed coordinate.  In stochastic energetics,
\begin{equation}
\dot Q_x=\int\dd x\,\dd\theta\,j_x[-V'(q)-f].
\label{eq:Qx_def}
\end{equation}
Solving Eq.~\eqref{eq:nux_raw} for the force and substituting into Eq.~\eqref{eq:Qx_def} gives
\begin{equation}
\frac{\dot Q_x}{T}=\Phi_x+\mathcal I_x,
\label{eq:Qx_split}
\end{equation}
where
\begin{equation}
\Phi_x=\int_0^L\rho(q)\frac{\nu_x(q)^2}{\mu_xT}\,\dd q
\label{eq:Phi_def}
\end{equation}
and
\begin{equation}
\mathcal I_x=\int_0^L\rho(q)\nu_x(q)\,\partial_q\ln\rho(q)\,\dd q.
\label{eq:I_def}
\end{equation}
The first term is the positive current-square contribution of the observed channel.  The second describes transport along gradients of the joint stationary density.  Because the $x$ and $\theta$ marginals are uniform under Eq.~\eqref{eq:factorization}, this correction coincides with the learning-rate or information-flow term used in bipartite stochastic thermodynamics~\cite{HorowitzEsposito2014}.  In a generic coupled system it does not vanish, so the heat measured through $x$ need not equal $T\Phi_x$.

For the reciprocal, translationally invariant motor, however, this correction vanishes at every steady state, so that $\dot Q_x=T\Phi_x$.  To see this, Eq.~\eqref{eq:nux_closure} gives
\begin{equation}
\nu_x(q)=\frac{1}{K}\left[\frac{J_q}{\rho(q)}
+\ell\mu_\theta(\tau_{\rm a}-\ell f)\right],
\label{eq:nux_explicit}
\end{equation}
and using this expression in Eq.~\eqref{eq:I_def},
\begin{align}
\mathcal I_x
&=\frac{1}{K}\int_0^L\left[\frac{J_q}{\rho}
+\ell\mu_\theta(\tau_{\rm a}-\ell f)\right]\rho'\,\dd q\nonumber\\
&=\frac{J_q}{K}\left[\ln\rho\right]_0^L
+\frac{\ell\mu_\theta(\tau_{\rm a}-\ell f)}{K}
\left[\rho\right]_0^L=0,
\label{eq:I_vanishes}
\end{align}
by periodicity and positivity of the finite-temperature density.  Equation~\eqref{eq:Qx_split} therefore reduces to $\dot Q_x=T\Phi_x$.

\paragraph*{Fluctuation--response measurement.}
Let $v(t)=\dot x(t)$ be the unwrapped observed velocity.  Define the connected spectrum and the response to a small probe force $h(t)$ added to Eq.~\eqref{eq:langevin_x} by
\begin{align}
C_{vv}(\omega)&=\int_{-\infty}^{\infty}\dd t\,\e^{i\omega t}
\avg{[v(t)-J_x][v(0)-J_x]},
\label{eq:Cvv}\\
R_v(\omega)&=\left.\frac{\delta\avg{v(\omega)}}{\delta h(\omega)}\right|_{h=0}.
\label{eq:Rv}
\end{align}
At equilibrium $C_{vv}=2T\,\mathrm{Re}\,R_v$.  Away from equilibrium, the Harada--Sasa equality~\cite{HaradaSasa2005,HaradaSasa2006} gives
\begin{equation}
\frac{\dot Q_x}{T}=\frac{J_x^2}{\mu_xT}
+\frac{1}{\mu_xT}\int_{-\infty}^{\infty}\frac{\dd\omega}{2\pi}
\left[C_{vv}(\omega)-2T\,\mathrm{Re}\,R_v(\omega)\right].
\label{eq:HS_general}
\end{equation}
The overdamped velocity contains white noise, so the two spectral terms separately possess the usual high-frequency contribution.  Equation~\eqref{eq:HS_general} is understood with the standard Harada--Sasa ultraviolet regularization; their difference has a finite integral.  At stall $J_x=0$, and the vanishing of the correlation-flow term shown above identifies the measured quantity with Eq.~\eqref{eq:Phi_def}:
\begin{equation}
\Phi_x^{\st}=\frac{1}{\mu_xT}\int_{-\infty}^{\infty}\frac{\dd\omega}{2\pi}
\left[C_{vv}(\omega)-2T\,\mathrm{Re}\,R_v(\omega)\right].
\label{eq:Phi_HS}
\end{equation}
Thus the stalled coordinate carries no information in its mean velocity, but its fluctuations and response measure the positive local traffic that remains in the observed channel.

\paragraph*{Decomposition of the full entropy production at stall.}
Averaging Eq.~\eqref{eq:local_identity} at stall, where $\avg{\nu_x}=J_x=0$, reproduces Eq.~\eqref{eq:hidden_stall_current}.  Pointwise,
\begin{equation}
\nu_\theta(q)=J_\theta^{\st}-\frac{\ell\mu_\theta}{\mu_x}\nu_x(q).
\label{eq:stall_local_identity}
\end{equation}
Squaring, averaging with $\rho$, and using $\avg{\nu_x}=0$ gives
\begin{equation}
\avg{\nu_\theta^2}=(J_\theta^{\st})^2
+\frac{\ell^2\mu_\theta^2}{\mu_x^2}\avg{\nu_x^2}.
\label{eq:nuth_square}
\end{equation}
Substitution into Eq.~\eqref{eq:ep} yields
\begin{equation}
\ep^{\st}=K\Phi_x^{\st}+\frac{(J_\theta^{\st})^2}{\mu_\theta T}.
\label{eq:stall_decomp}
\end{equation}
The first term is not simply the visible dissipation $\Phi_x^{\st}$: it is amplified by $K$ because each local fluctuation of $x$ induces reciprocal hidden motion.  The second term is the dissipation associated with the nonzero mean rotation of the hidden cycle.

At stall, the power balance \eqref{eq:stall_ep} and Eq.~\eqref{eq:hidden_stall_current} imply
\begin{equation}
\tau_{\rm a}=\ell f_{\st}+\frac{J_\theta^{\st}}{\mu_\theta}
\end{equation}
and therefore
\begin{equation}
T\ep^{\st}=\ell f_{\st}J_\theta^{\st}
+\frac{(J_\theta^{\st})^2}{\mu_\theta}.
\label{eq:power_expanded}
\end{equation}
Multiplying Eq.~\eqref{eq:stall_decomp} by $T$ and comparing it with Eq.~\eqref{eq:power_expanded} cancels the same quadratic hidden-current term on both sides.  The remaining relation is
\begin{equation}
TK\Phi_x^{\st}=\ell f_{\st}J_\theta^{\st}.
\label{eq:inference_identity}
\end{equation}
This identity is the central closure relation: the hidden current is fixed by the visible fluctuation--response violation and the reciprocal stall load.

This closure relation yields the exact stall-point reconstruction.  For the reciprocal hidden-rotor motor with $\ell f_{\st}\neq0$, if $K=1+\ell^2\mu_\theta/\mu_x$ is known, then Eq.~\eqref{eq:inference_identity} gives directly
\begin{equation}
J_\theta^{\st}=\frac{TK\Phi_x^{\st}}{\ell f_{\st}},
\label{eq:Jth_reconstructed}
\end{equation}
and substituting this into the decomposition~\eqref{eq:stall_decomp} gives
\begin{equation}
\ep^{\st}=K\Phi_x^{\st}
+\frac{TK^2(\Phi_x^{\st})^2}{\mu_\theta\ell^2 f_{\st}^2}.
\label{eq:recon_intermediate}
\end{equation}
Since $\ell^2\mu_\theta=\mu_x(K-1)$, the total entropy production is exactly
\begin{equation}
\ep^{\st}=K\Phi_x^{\st}
+\frac{TK^2(\Phi_x^{\st})^2}{\mu_x(K-1)f_{\st}^2}.
\label{eq:exact_reconstruction}
\end{equation}

The inputs are $f_{\st}$, $T$, $\mu_x$, $\ell$, and the measured spectrum and response entering $\Phi_x^{\st}$, together with $K$.  The hidden mobility in $K$ is not determined by $x$ alone and must be calibrated independently, for example by transiently resolving the rotor, weakening the coupling and measuring its torque response, or using a separate measurement of the internal relaxation time.

\paragraph*{Unknown hidden mobility: optimized reciprocal bound.}
Even without such calibration, reciprocity restricts $K>1$.  Define the dimensionless measured ratio
\begin{equation}
a=\frac{T\Phi_x^{\st}}{\mu_xf_{\st}^2}.
\label{eq:a}
\end{equation}
Dividing Eq.~\eqref{eq:exact_reconstruction} by $\Phi_x^{\st}$ gives
\begin{equation}
\frac{\ep^{\st}}{\Phi_x^{\st}}
=K+a\frac{K^2}{K-1}.
\label{eq:K_function}
\end{equation}
Writing $K=1+s$, $s>0$, the right-hand side becomes
\begin{equation}
1+2a+(1+a)s+\frac{a}{s}.
\end{equation}
Its derivative vanishes at
\begin{equation}
s_\ast=\sqrt{\frac{a}{1+a}},
\qquad
K_\ast=1+\sqrt{\frac{a}{1+a}},
\label{eq:Kstar}
\end{equation}
and the minimum is $(\sqrt{1+a}+\sqrt a)^2$.  Hence, for an unknown $K>1$ and $\ell f_{\st}\neq0$,
\begin{equation}
\ep^{\st}\geq\Phi_x^{\st}\left(\sqrt{1+a}+\sqrt a\right)^2,
\qquad
a=\frac{T\Phi_x^{\st}}{\mu_xf_{\st}^2},
\label{eq:tight_bound}
\end{equation}
a bound that, as we now show, is attained by a member of the reciprocal hidden-rotor class.

The factor multiplying $\Phi_x^{\st}$ is greater than unity for any nonzero visible dissipation.  The elementary channel bound $\ep^{\st}\geq\Phi_x^{\st}$ discards all information about reciprocal back reaction; Eq.~\eqref{eq:tight_bound} uses the fact that observed local motion necessarily drags the hidden coordinate.

To establish attainability rather than only algebraic optimality, fix target values $f_{\st}>0$ and $\Phi_x^{\st}>0$, choose $K=K_\ast$, and set
\begin{equation}
\mu_\theta=\frac{\mu_x(K_\ast-1)}{\ell^2},
\quad
J_\theta^{\st}=\frac{TK_\ast\Phi_x^{\st}}{\ell f_{\st}},
\quad
\tau_{\rm a}=\ell f_{\st}+\frac{J_\theta^{\st}}{\mu_\theta}.
\label{eq:saturation_parameters}
\end{equation}
Consider $V_A(q)=A\cos(2\pi q/L)$.  At $A=0$,
\begin{equation}
v_q(0)=-(\mu_xf_{\st}+\ell\mu_\theta\tau_{\rm a})
<-\ell J_\theta^{\st},
\end{equation}
whereas $v_q(A)\to0>-\ell J_\theta^{\st}$ as $A\to\infty$.  The tilted-periodic current varies continuously with $A$, so an amplitude $A_\ast$ exists for which $v_q(A_\ast)=-\ell J_\theta^{\st}$, exactly the stall condition.  Equation~\eqref{eq:inference_identity} then fixes the target $\Phi_x^{\st}$.  The lower bound is therefore reachable within the physical model class.

\paragraph*{Observability limit and protocol.}
The factor $\ell f_{\st}$ in Eq.~\eqref{eq:inference_identity} expresses a genuine observability condition.  If $\ell=0$, the hidden rotor is mechanically decoupled and can dissipate without affecting $x$.  If $f_{\st}=0$, the measured stall load carries no reciprocal leverage with which to infer the hidden current.  In either case, $x$ alone cannot reconstruct arbitrary hidden dissipation.  In the observable regime, an experiment locates $f_{\st}$ from the zero of $J_x$, measures $\mu_x$, records the velocity fluctuations and linear response at stall to obtain $\Phi_x^{\st}$, and then uses Eq.~\eqref{eq:exact_reconstruction} when $K$ is calibrated or Eq.~\eqref{eq:tight_bound} when it is not.

\section{Analytical regimes and numerical verification}
\label{sec:verification}

The exact formulas do not require a special potential.  For numerical illustration we use
\begin{equation}
V(q)=V_0\cos(2\pi q/L)
\label{eq:sin_potential}
\end{equation}
and dimensionless units $L=T=\mu_x=1$, with $\ell=L/(2\pi)$.  For each parameter set, Eq.~\eqref{eq:Jq_exact} is evaluated by direct quadrature, the root of Eq.~\eqref{eq:stall_condition} determines $f_{\st}$, and Eq.~\eqref{eq:rho_exact} determines the stationary density.  The local velocities are evaluated from Eqs.~\eqref{eq:nux_explicit} and \eqref{eq:local_identity}, avoiding numerical differentiation, and the direct entropy production follows from Eq.~\eqref{eq:ep}.  The same parameters are also integrated in the original two-variable Langevin equations, without using the reduced coordinate in the simulation.

The computations verify four logically distinct identities.  First, the root satisfies $J_x(f_{\st})=0$ while $J_\theta^{\st}$ remains finite.  Second, the direct current-square entropy production equals the power input, $T\ep^{\st}=\tau_{\rm a}J_\theta^{\st}$.  Third, the correlation-flow term in Eq.~\eqref{eq:I_def} converges to zero.  Fourth, the total entropy production from the full current field coincides with Eq.~\eqref{eq:exact_reconstruction}.  These tests separately check the transport reduction, thermodynamic balance, observable-channel identification, and inference theorem.

Over a range of coupling strengths $V_0$, with $\tau_{\rm a}=1$ and $\mu_\theta=\mu_x$, the direct stall dissipation remains close to the chemical input, whereas the visible contribution $\Phi_x^{\st}$ alone is much smaller.  The optimized reciprocal bound uses the measured stall load to recover part of the hidden cost and therefore lies strictly above the visible bound, while the exact reconstruction of Eq.~\eqref{eq:exact_reconstruction} coincides with the direct result.  The equality persists when the hidden mobility is varied over $\mu_\theta/\mu_x=0.5,1,2$, a nontrivial test because both the gearing factor $K$ and the hidden-current contribution change with $\mu_\theta$.

Representative trajectory simulations give the same stall currents.  Euler--Maruyama trajectories were divided into long blocks after discarding an initial transient, and the quoted uncertainties are standard errors of the block currents.  At $(V_0,\tau_{\rm a},\mu_\theta)=(0.6,1,1)$, quadrature gives $f_{\st}=0.02972$ and $J_\theta^{\st}=0.9953$, while direct trajectories give $J_x=-0.0036\pm0.0072$ and $J_\theta=0.9982\pm0.0081$.  At $(1.0,1,2)$, quadrature gives $f_{\st}=0.17650$ and $J_\theta^{\st}=1.9438$, while trajectories give $J_x=-0.0034\pm0.0070$ and $J_\theta=1.9317\pm0.0119$.  The numerical agreement does not establish the theorem---which follows analytically---but it checks that no reduction, sign, or normalization convention has been used inconsistently.

Several limiting regimes clarify the result.  In the equilibrium limit $f=\tau_{\rm a}=0$, all currents and the fluctuation--response violation vanish.  In the uncoupled limit $V_0\to0$, the stall load and $\Phi_x^{\st}$ vanish even though the isolated rotor can dissipate; this is precisely the unobservable case identified above.  In the strong-coupling limit, slip is suppressed and the translation follows the rotor more tightly, increasing the mechanical stall load.  The inference formula remains exact throughout because it depends on reciprocity and symmetry rather than on the coupling strength.

\section{Energetics, power conversion, and operating regimes}
\label{sec:energetics}

The reciprocal hidden-rotor model is an energy-conversion system with two external generalized forces and two dissipative channels. The active torque $\tau_{\rm a}$ acts on the hidden angular coordinate $\theta$, while the mechanical load $f$ acts against translation along $x$. The periodic interaction
\begin{equation}
U(x,\theta)=V(q),
\qquad
q=x-\ell\theta ,
\label{eq:energy_potential}
\end{equation}
is conservative. It stores energy temporarily and transfers energy between the two coordinates, but it does not itself supply energy. The only nonconservative agents are the active torque and the external load.

Throughout this section, heat is defined as positive when energy is released by the system into the thermal environment. Work is positive when it is supplied to the system unless explicitly identified as useful output. Because the Langevin noise amplitudes are constant, the It\^o and Stratonovich conventions generate the same state dynamics. Stochastic work and heat increments are nevertheless written in the Stratonovich convention, denoted by $\circ$, because this convention preserves the ordinary chain rule of stochastic energetics~\cite{Sekimoto2010,Seifert2012}.

Although $x$ and $\theta$ are periodic variables, their currents and accumulated work are defined on their unwrapped lifts. Thus $J_x$ counts the net number of spatial periods traversed per unit time, while $J_\theta$ counts the net angular winding rate.

\paragraph*{Trajectory-level first law.}

The conservative force on the translational coordinate and the reciprocal torque on the hidden coordinate are
\begin{equation}
F_x^{\rm int}
=
-\partial_x U
=
-V'(q),
\qquad
\tau_\theta^{\rm int}
=
-\partial_\theta U
=
\ell V'(q),
\label{eq:internal_force_torque}
\end{equation}
and consequently
\begin{equation}
\tau_\theta^{\rm int}
=
-\ell F_x^{\rm int}.
\label{eq:reciprocity_energy}
\end{equation}
The total deterministic generalized forces appearing in the Langevin equations are therefore
\begin{equation}
F_x^{\rm tot}
=
-V'(q)-f,
\qquad
\tau_\theta^{\rm tot}
=
\tau_{\rm a}+\ell V'(q).
\label{eq:total_forces}
\end{equation}

The active torque performs the infinitesimal work
\begin{equation}
\delta W_a
=
\tau_{\rm a}\circ d\theta ,
\label{eq:active_work_increment}
\end{equation}
whereas the load performs work on the system according to
\begin{equation}
\delta W_{\rm load}^{\rm on}
=
-f\circ dx .
\label{eq:load_work_on}
\end{equation}
It is convenient to define the useful mechanical work delivered by the system to the load as
\begin{equation}
\delta W_{\rm out}
=
-\delta W_{\rm load}^{\rm on}
=
f\circ dx .
\label{eq:output_work_increment}
\end{equation}

The heat increments released through the two overdamped bath channels are
\begin{align}
\delta Q_x
&=
\bigl[-V'(q)-f\bigr]\circ dx ,
\label{eq:heat_x_increment}
\\
\delta Q_\theta
&=
\bigl[\tau_{\rm a}+\ell V'(q)\bigr]\circ d\theta .
\label{eq:heat_theta_increment}
\end{align}
Using
\begin{equation}
dU
=
V'(q)\circ dq
=
V'(q)\circ dx
-
\ell V'(q)\circ d\theta ,
\label{eq:dU_chain_rule}
\end{equation}
one obtains the trajectory-level first law
\begin{equation}
dU
=
\delta W_a
-
\delta W_{\rm out}
-
\delta Q_x
-
\delta Q_\theta
.
\label{eq:trajectory_first_law}
\end{equation}
Equation~\eqref{eq:trajectory_first_law} has a direct interpretation. The active agent supplies energy through rotation of the hidden coordinate. Part of this energy can be delivered as mechanical work against the load, part can be stored transiently in the coupling potential, and the remainder is released as heat through the $x$ and $\theta$ bath channels.

The interaction potential is not counted as an input source. Its contributions cancel between the two coordinates because the force and torque derive from the same scalar potential. This cancellation is the energetic content of the reciprocal relation~\eqref{eq:reciprocity_energy}.

\paragraph*{Ensemble first law and signed powers.}

The mean internal energy is
\begin{equation}
\mathcal U(t)
=
\langle U\rangle
=
\int dx\,d\theta\,
p(x,\theta,t)U(x,\theta).
\label{eq:mean_internal_energy}
\end{equation}
Using the continuity equation and integrating by parts on the torus,
\begin{equation}
\frac{d\mathcal U}{dt}
=
\int dx\,d\theta
\left[
\partial_xU\,j_x
+
\partial_\theta U\,j_\theta
\right].
\label{eq:mean_energy_derivative}
\end{equation}

We define the signed active power
\begin{equation}
P_a
=
\tau_aJ_\theta ,
\label{eq:active_power}
\end{equation}
the signed power supplied to the system by the mechanical load
\begin{equation}
P_{\rm load}^{\rm on}
=
-fJ_x ,
\label{eq:load_power_on}
\end{equation}
and the mechanical output power
\begin{equation}
P_{\rm out}
=
-P_{\rm load}^{\rm on}
=
fJ_x .
\label{eq:output_power}
\end{equation}
The mean heat-release rates are
\begin{align}
\dot Q_x
&=
\int dx\,d\theta\,
\bigl[-V'(q)-f\bigr]j_x,
\label{eq:mean_heat_x}
\\
\dot Q_\theta
&=
\int dx\,d\theta\,
\bigl[\tau_{\rm a}+\ell V'(q)\bigr]j_\theta .
\label{eq:mean_heat_theta}
\end{align}
Averaging Eq.~\eqref{eq:trajectory_first_law} gives
\begin{equation}
\frac{d\mathcal U}{dt}
=
P_a
-
P_{\rm out}
-
\dot Q_x
-
\dot Q_\theta
.
\label{eq:ensemble_first_law}
\end{equation}

The quantities in Eq.~\eqref{eq:ensemble_first_law} are signed. In particular, $P_a>0$ means that the active torque supplies energy to the system, whereas $P_a<0$ means that the system performs work on the active driving reservoir. Similarly, $P_{\rm out}>0$ means that the motor performs useful work on the load, whereas $P_{\rm out}<0$ means that the mechanical load supplies energy to the system.

The term ``input power'' should therefore be used only after the operating regime has been identified. In the forward motor regime,
\begin{equation}
P_a>0,
\qquad
P_{\rm out}>0,
\label{eq:motor_regime_signs}
\end{equation}
and the active power is the input:
\begin{equation}
P_{\rm in}=P_a=\tau_aJ_\theta .
\label{eq:input_power_motor}
\end{equation}
Outside this regime, $\tau_aJ_\theta$ remains the signed active power but need not be the system's input.

\paragraph*{Chemical interpretation of the active torque.}

The Langevin model itself contains a generic nonconservative torque $\tau_{\rm a}$. Calling this torque ``chemical'' requires a physical identification of $\theta$ with progress around a chemical reaction cycle.

Suppose that one complete angular revolution,
\begin{equation}
\Delta\theta=2\pi ,
\end{equation}
corresponds to one chemical turnover with available free-energy drop $\Delta G_{\rm cyc}$. A constant coarse-grained torque representing this affinity is
\begin{equation}
\tau_{\rm a}
=
\frac{\Delta G_{\rm cyc}}{2\pi}.
\label{eq:chemical_torque_mapping}
\end{equation}
The chemical turnover rate is
\begin{equation}
r_{\rm cyc}
=
\frac{J_\theta}{2\pi},
\label{eq:turnover_rate}
\end{equation}
and hence the chemical free-energy consumption rate is
\begin{equation}
P_{\rm chem}
=
\Delta G_{\rm cyc}\,r_{\rm cyc}
=
\frac{\Delta G_{\rm cyc}}{2\pi}J_\theta
=
\tau_aJ_\theta .
\label{eq:chemical_power}
\end{equation}
Thus
\begin{equation}
P_{\rm chem}=P_a=\tau_aJ_\theta
\label{eq:chemical_active_equivalence}
\end{equation}
when $\theta$ is a tightly coupled chemical-cycle coordinate.

If no explicit chemical interpretation is intended, the more general terminology
\begin{equation}
P_a=\tau_aJ_\theta
\end{equation}
should be used, and it should be called the active or nonconservative driving power rather than chemical power.

\paragraph*{Entropy balance and nonequilibrium free energy.}

The Shannon entropy of the two-coordinate system is
\begin{equation}
S(t)
=
-\int dx\,d\theta\,
p(x,\theta,t)\ln p(x,\theta,t).
\label{eq:shannon_energetics}
\end{equation}
The total entropy-production rate is
\begin{equation}
\ep
=
\int dx\,d\theta
\left[
\frac{j_x^2}{\mu_xTp}
+
\frac{j_\theta^2}{\mu_\theta Tp}
\right]
\geq0.
\label{eq:epr_energetics}
\end{equation}
With heat defined as positive into the environment, the entropy balance is
\begin{equation}
\ep
=
\frac{dS}{dt}
+
\frac{\dot Q_x+\dot Q_\theta}{T}
.
\label{eq:entropy_balance_energetics}
\end{equation}
Equivalently, if $\hdr$ denotes entropy extraction from the system,
\begin{equation}
\hdr
=
\frac{\dot Q_x+\dot Q_\theta}{T},
\qquad
\frac{dS}{dt}
=
\ep-\hdr.
\label{eq:entropy_extraction_heat}
\end{equation}

Define the nonequilibrium free energy
\begin{equation}
\mathcal F(t)
=
\mathcal U(t)-TS(t).
\label{eq:noneq_free_energy}
\end{equation}
Combining the first law~\eqref{eq:ensemble_first_law} with the entropy balance~\eqref{eq:entropy_balance_energetics} gives
\begin{equation}
\frac{d\mathcal F}{dt}
=
P_a-P_{\rm out}-T\ep
.
\label{eq:free_energy_balance}
\end{equation}
Equation~\eqref{eq:free_energy_balance} separates the three possible destinations of supplied power: useful mechanical output, irreversible dissipation, and transient free-energy storage.

At a steady state,
\begin{equation}
\frac{d\mathcal U}{dt}
=
\frac{dS}{dt}
=
\frac{d\mathcal F}{dt}
=
0,
\end{equation}
and therefore
\begin{equation}
T\ep
=
P_a-P_{\rm out}
=
\tau_aJ_\theta-fJ_x
.
\label{eq:steady_power_balance_detailed}
\end{equation}
This identity is the steady first and second laws combined. It states that the difference between the signed active power and the mechanical output is dissipated as heat.

In the forward motor regime,
\begin{equation}
P_{\rm in}=P_a,
\end{equation}
so Eq.~\eqref{eq:steady_power_balance_detailed} becomes
\begin{equation}
P_{\rm in}
=
P_{\rm out}
+
T\ep.
\label{eq:motor_energy_balance}
\end{equation}

\paragraph*{Internal transfer of energy from the rotor to translation.}

The conservative interaction transfers energy between the hidden rotor and the observed coordinate. Define the mean power transferred from the rotor sector to the translational sector by
\begin{equation}
P_{\theta\rightarrow x}
=
-\int dx\,d\theta\,
V'(q)j_x.
\label{eq:internal_transfer_power}
\end{equation}
Positive $P_{\theta\rightarrow x}$ means that the coupling delivers energy to the observed coordinate.

At steady state,
\begin{equation}
0
=
\frac{d\mathcal U}{dt}
=
\int dx\,d\theta\,
\left[
V'(q)j_x
-
\ell V'(q)j_\theta
\right],
\label{eq:steady_conservative_balance}
\end{equation}
and hence
\begin{equation}
P_{\theta\rightarrow x}
=
-\ell
\int dx\,d\theta\,
V'(q)j_\theta .
\label{eq:transfer_reciprocity}
\end{equation}
The two channel balances can therefore be written as
\begin{align}
P_a
&=
\dot Q_\theta
+
P_{\theta\rightarrow x},
\label{eq:rotor_channel_balance}
\\
P_{\theta\rightarrow x}
&=
P_{\rm out}
+
\dot Q_x.
\label{eq:translation_channel_balance}
\end{align}
Equations~\eqref{eq:rotor_channel_balance} and
\eqref{eq:translation_channel_balance} give the energetic pathway
\begin{equation}
P_a
\longrightarrow
P_{\theta\rightarrow x}
\longrightarrow
P_{\rm out},
\end{equation}
with heat released both before and after the internal transfer:
\begin{equation}
P_a
=
\dot Q_\theta
+
\dot Q_x
+
P_{\rm out}.
\label{eq:energetic_circuit}
\end{equation}
The coupling potential acts as a transmission element. It carries power from the hidden rotor to the observed coordinate but contributes no net power of its own.

\paragraph*{Channel-resolved dissipation.}

Let
\begin{equation}
\nu_x=\frac{j_x}{p},
\qquad
\nu_\theta=\frac{j_\theta}{p}
\label{eq:local_velocities_energy}
\end{equation}
be the local mean velocities. Define the positive current-square contributions
\begin{align}
\Phi_x
&=
\int dx\,d\theta\,
p\frac{\nu_x^2}{\mu_xT},
\label{eq:phi_x_energy}
\\
\Phi_\theta
&=
\int dx\,d\theta\,
p\frac{\nu_\theta^2}{\mu_\theta T}.
\label{eq:phi_theta_energy}
\end{align}
Then
\begin{equation}
\ep
=
\Phi_x+\Phi_\theta .
\label{eq:epr_channel_sum}
\end{equation}

For a generic coupled bipartite system, the heat released through one coordinate need not equal the corresponding positive current-square term. Instead,
\begin{align}
\frac{\dot Q_x}{T}
&=
\Phi_x+I_x,
&
I_x
&=
\int dx\,d\theta\,
p\nu_x\partial_x\ln p,
\label{eq:heat_phi_x_general}
\\
\frac{\dot Q_\theta}{T}
&=
\Phi_\theta+I_\theta,
&
I_\theta
&=
\int dx\,d\theta\,
p\nu_\theta\partial_\theta\ln p.
\label{eq:heat_phi_theta_general}
\end{align}
The terms $I_x$ and $I_\theta$ represent correlation or information-flow corrections between the two coordinates~\cite{HorowitzEsposito2014}.

For the present reciprocal and translationally invariant model,
\begin{equation}
p_{\rm ss}(x,\theta)
=
\frac{\rho(q)}{2\pi},
\end{equation}
and the local current identity gives
\begin{equation}
\nu_x(q)
=
\frac{1}{K}
\left[
\frac{J_q}{\rho(q)}
+
\ell\mu_\theta
\bigl(\tau_{\rm a}-\ell f\bigr)
\right],
\qquad
K=1+\frac{\ell^2\mu_\theta}{\mu_x}.
\label{eq:nux_energy_identity}
\end{equation}
Consequently,
\begin{align}
I_x
&=
\int_0^L dq\,
\nu_x(q)\rho'(q)
\nonumber\\
&=
\frac{1}{K}
\left[
J_q\int_0^L
\frac{\rho'(q)}{\rho(q)}\,dq
+
\ell\mu_\theta(\tau_{\rm a}-\ell f)
\int_0^L\rho'(q)\,dq
\right]
=0
\label{eq:Ix_energy_zero}
\end{align}
by periodicity. The reciprocal local relation
\begin{equation}
\nu_\theta(q)
=
\mu_\theta(\tau_{\rm a}-\ell f)
-
\frac{\ell\mu_\theta}{\mu_x}\nu_x(q)
\label{eq:nutheta_energy_identity}
\end{equation}
then also gives
\begin{equation}
I_\theta
=
-\ell\int_0^L
\nu_\theta(q)\rho'(q)\,dq
=
0.
\label{eq:Itheta_energy_zero}
\end{equation}
Hence, in this particular steady state,
\begin{equation}
\dot Q_x=T\Phi_x,
\qquad
\dot Q_\theta=T\Phi_\theta
.
\label{eq:channel_heat_equals_phi}
\end{equation}
The two bath channels therefore have separate positive dissipation rates. This is a special consequence of reciprocal coupling and translational symmetry; it is not generally true for arbitrary partially observed systems.

The first equality in Eq.~\eqref{eq:channel_heat_equals_phi} is also the reason that the Harada--Sasa fluctuation--response violation of the observed coordinate has a direct energetic meaning:
\begin{equation}
\dot Q_x
=
\frac{J_x^2}{\mu_x}
+
\frac{1}{\mu_x}
\int_{-\infty}^{\infty}
\frac{d\omega}{2\pi}
\left[
C_{vv}(\omega)
-
2T\,{\rm Re}\,R_v(\omega)
\right].
\label{eq:harada_sasa_energetics}
\end{equation}

\paragraph*{Energetics at mechanical stall.}

Mechanical stall is defined by
\begin{equation}
J_x(f_{\st})=0.
\label{eq:stall_energy_definition}
\end{equation}
Therefore
\begin{equation}
P_{\rm out}^{\st}
=
f_{\st}J_x(f_{\st})
=
0.
\label{eq:stall_output_zero}
\end{equation}
The hidden angular current remains
\begin{equation}
J_\theta^{\st}
=
\mu_\theta
\bigl(\tau_{\rm a}-\ell f_{\st}\bigr).
\label{eq:stall_hidden_current_energy}
\end{equation}
On the forward motor branch,
\begin{equation}
\tau_aJ_\theta^{\st}>0,
\end{equation}
the active torque continues to supply energy even though the observed coordinate produces no average output. The steady power balance becomes
\begin{equation}
P_{\rm in}^{\st}
=
P_a^{\st}
=
\tau_aJ_\theta^{\st}
=
T\ep^{\st}
=
\dot Q_x^{\st}
+
\dot Q_\theta^{\st}
.
\label{eq:stall_complete_energy_balance}
\end{equation}
Thus all of the input power is converted into heat at stall.

The internal transfer relation becomes particularly transparent:
\begin{equation}
P_{\theta\rightarrow x}^{\st}
=
\dot Q_x^{\st}
=
T\Phi_x^{\st}.
\label{eq:stall_transfer_visible_heat}
\end{equation}
Even though the mean translational current is zero, the rotor continuously transfers energy into the $x$ coordinate. That energy drives phase-dependent forward and backward motions whose signed displacements cancel but whose heat production does not.

Using the reciprocal local identity at stall,
\begin{equation}
\nu_\theta(q)
=
J_\theta^{\st}
-
\frac{\ell\mu_\theta}{\mu_x}\nu_x(q),
\end{equation}
the hidden-channel dissipation is
\begin{equation}
\Phi_\theta^{\st}
=
(K-1)\Phi_x^{\st}
+
\frac{(J_\theta^{\st})^2}{\mu_\theta T}.
\label{eq:hidden_channel_stall_dissipation}
\end{equation}
Therefore
\begin{align}
\dot Q_x^{\st}
&=
T\Phi_x^{\st},
\label{eq:stall_visible_heat}
\\
\dot Q_\theta^{\st}
&=
T(K-1)\Phi_x^{\st}
+
\frac{(J_\theta^{\st})^2}{\mu_\theta},
\label{eq:stall_hidden_heat}
\\
T\ep^{\st}
&=
TK\Phi_x^{\st}
+
\frac{(J_\theta^{\st})^2}{\mu_\theta}.
\label{eq:stall_total_heat_channels}
\end{align}
The first contribution to Eq.~\eqref{eq:stall_hidden_heat} is the additional hidden dissipation forced by reciprocal fluctuations. The second is the direct frictional cost of the nonzero mean angular current.

When $K$ is calibrated, the hidden current is reconstructed from the observed stall-point fluctuation--response violation:
\begin{equation}
J_\theta^{\st}
=
\frac{TK\Phi_x^{\st}}
{\ell f_{\st}}.
\label{eq:hidden_current_energy_reconstruction}
\end{equation}
The total input power at stall is consequently
\begin{equation}
P_{\rm in}^{\st}
=
TK\Phi_x^{\st}
+
\frac{
T^2K^2(\Phi_x^{\st})^2
}{
\mu_x(K-1)f_{\st}^2
}
.
\label{eq:stall_input_reconstruction}
\end{equation}
The hidden-channel heat is
\begin{equation}
\dot Q_\theta^{\st}
=
T(K-1)\Phi_x^{\st}
+
\frac{
T^2K^2(\Phi_x^{\st})^2
}{
\mu_x(K-1)f_{\st}^2
}
.
\label{eq:hidden_heat_reconstruction}
\end{equation}
Thus the reconstruction theorem can be read not only as an entropy-production result, but also as a complete energetic reconstruction: it determines the total fuel or active power consumed at stall and how that power is divided between visible and hidden heat.

If $K$ is unknown, the optimized reciprocal bound gives the energetic lower bound
\begin{equation}
P_{\rm in}^{\st}
=
T\ep^{\st}
\geq
T\Phi_x^{\st}
\left(
\sqrt{1+a}+\sqrt{a}
\right)^2
,
\qquad
a=
\frac{T\Phi_x^{\st}}
{\mu_x f_{\st}^2}.
\label{eq:stall_input_bound}
\end{equation}

\paragraph*{Efficiency and reverse operation.}

In the forward motor regime, where
\begin{equation}
P_a>0,
\qquad
P_{\rm out}>0,
\end{equation}
the conversion efficiency is
\begin{equation}
\eta
=
\frac{P_{\rm out}}{P_{\rm in}}
=
\frac{fJ_x}{\tau_aJ_\theta}.
\label{eq:motor_efficiency_energy}
\end{equation}
The steady power balance gives
\begin{equation}
\eta
=
1-
\frac{T\ep}{\tau_aJ_\theta},
\label{eq:efficiency_dissipation}
\end{equation}
and therefore
\begin{equation}
0\leq\eta\leq1.
\label{eq:efficiency_bound}
\end{equation}

At mechanical stall,
\begin{equation}
\eta_{\st}=0
\label{eq:stall_efficiency_zero}
\end{equation}
whenever $\tau_aJ_\theta^{\st}>0$, because the motor consumes positive input power while producing no mechanical output. Stall is therefore not a high-efficiency reversible state. It is a zero-output, finite-consumption nonequilibrium state.

If $J_x<0$ for $f>0$, then
\begin{equation}
P_{\rm out}=fJ_x<0,
\end{equation}
and the load supplies mechanical power
\begin{equation}
P_{\rm mech,in}
=
-P_{\rm out}
=
-fJ_x>0.
\label{eq:mechanical_input_reverse}
\end{equation}
If at the same time $P_a=\tau_aJ_\theta<0$, the device converts mechanical input into work delivered to the active or chemical reservoir. A reverse conversion efficiency may then be defined by
\begin{equation}
\eta_{\rm rev}
=
\frac{-P_a}{P_{\rm mech,in}}
=
\frac{-\tau_aJ_\theta}{-fJ_x},
\qquad
0\leq\eta_{\rm rev}\leq1.
\label{eq:reverse_efficiency}
\end{equation}
If both $P_a>0$ and $P_{\rm out}<0$, both external agents supply energy and the system acts as a driven dissipator rather than an energy converter.

The word ``input'' is therefore regime dependent. In the forward motor regime and at its mechanical stall point,
\begin{equation}
P_{\rm in}=\tau_aJ_\theta.
\end{equation}
In reverse operation, the mechanical load may instead be the input source. The signed balance
\begin{equation}
T\ep
=
\tau_aJ_\theta-fJ_x
\end{equation}
remains valid in every regime and determines which agent supplies energy from the signs of its associated power.

\section{Discussion and conclusions}
\label{sec:discussion}

The central conceptual distinction in this work is between a constraint imposed on an observed degree of freedom and reversibility of the full stochastic dynamics. Mechanical stall imposes only the averaged condition $J_x=0$ on the measured coordinate. Thermodynamic reversibility, by contrast, requires the complete steady-state probability-current field to vanish locally, $j_x(x,\theta)=j_\theta(x,\theta)=0$. The former may arise from cancellations in the observed motion, whereas the latter admits no such cancellation because entropy production is a positive quadratic functional of the full current field. In the reciprocal motor studied here, the observed coordinate may therefore be mechanically stalled while the hidden rotor continues to circulate. This residual motion appears as a nonzero current of the relative slip variable $q=x-\ell\theta$, consumes active input power, and produces entropy without generating mechanical output.

The model is intentionally minimal, but its minimality is useful.  A shared potential makes the reaction torque explicit and removes ambiguity about where energy enters and leaves.  The exact power balance then identifies the residual entropy production at stall as the active power consumed by the hidden rotor.  Translational symmetry makes the stationary state one-dimensional, while reciprocity ties the local hidden and visible velocities.  These two structural properties, rather than a particular form of $V$, are responsible for the inference result.

The exact reconstruction theorem has a specific domain of validity. It is not a general formula for all hidden-variable systems. It applies to an isothermal overdamped motor with one hidden reciprocal coordinate, where the coupling depends only on the relative phase $q=x-\ell\theta$. If the hidden dynamics is not mechanically tied to the observed coordinate, or if additional hidden coordinates affect the entropy balance, measurements of $x$ alone may not determine the total thermodynamic cost. This limitation is clear in the singular cases $\ell=0$ and $f_{\st}=0$: when $\ell=0$, the hidden coordinate is decoupled from $x$; when $f_{\st}=0$, the stall load provides no leverage for inferring the hidden current. Within the reciprocal class studied here, however, the reconstruction requires neither time-scale separation nor knowledge of the detailed form of the periodic potential.

The result goes beyond simply detecting nonequilibrium behavior. Measuring fluctuation--response violation in the stalled coordinate indicates whether the observed motion is equilibrium-like. However, such a measurement alone does not generally determine the total entropy production. In the reciprocal motor considered here, the situation is more constrained. The observed fluctuation--response violation, together with the stall load and the calibrated gearing mobility, determines the full stall dissipation exactly. If the hidden mobility is not calibrated, the same reciprocal structure still yields an optimized lower bound on the total dissipation. This bound is more restrictive than the visible-channel bound and is attainable within the reciprocal model class.

Possible realizations include molecular motors tracked by an attached bead, rotary-to-linear mechanochemical converters, and active colloids whose observed motion is coupled to an unresolved internal propulsion phase. In these systems, the hidden mobility factor could be calibrated separately, while the stall load, velocity spectrum, and response function are measured from the observed coordinate. Our theoretical framework can then be used to infer the active or chemical power dissipated by a motor that appears mechanically motionless on average.

In summary, the reciprocal hidden-rotor model turns a familiar fact into a quantitative inference principle: mechanical stall is not thermodynamic equilibrium. The steady state is solved exactly through the slip coordinate, which allows the entropy and power balances to be written in closed form and makes the true detailed-balance condition explicit. Most importantly, the hidden entropy production at stall can be reconstructed exactly when the reciprocal mobility factor is known, or bounded sharply from the noise and response of a single observed coordinate when it is not. Thus, a mechanically silent motor may remain thermodynamically active; under reciprocal coupling, this hidden activity is not only detectable but quantitatively inferable.

\appendix

\section{Detailed entropy and energy balances}
\label{app:entropy}

Starting from Eq.~\eqref{eq:entropy}, normalization gives
\begin{equation}
\dot S=-\int\dd x\,\dd\theta\,(\partial_t p)\ln p.
\end{equation}
Using Eq.~\eqref{eq:fpe} and integrating by parts on the torus,
\begin{equation}
\dot S=-\int\dd x\,\dd\theta\,
(j_x\partial_x\ln p+j_\theta\partial_\theta\ln p).
\end{equation}
For $i=x,\theta$, write $j_i=\mu_iF_i p-\mu_iT\partial_i p$, where
\begin{equation}
F_x=-V'(q)-f,
\qquad
F_\theta=\tau_{\rm a}+\ell V'(q).
\end{equation}
Then
\begin{equation}
\partial_i\ln p=\frac{F_i}{T}-\frac{j_i}{\mu_iTp},
\end{equation}
and therefore
\begin{equation}
\dot S=\sum_i\int\frac{j_i^2}{\mu_iTp}
-\frac{1}{T}\sum_i\int j_iF_i,
\end{equation}
which is Eqs.~\eqref{eq:entropy_balance}--\eqref{eq:hd}.

For the energy balance,
\begin{align}
\frac{\dd}{\dd t}\avg U
&=\int U\partial_t p
=-\int U(\partial_xj_x+\partial_\theta j_\theta)\nonumber\\
&=\int(\partial_xU\,j_x+\partial_\theta U\,j_\theta)
=\int V'(q)(j_x-\ell j_\theta).
\end{align}
Meanwhile,
\begin{align}
T\hdr
&=\int[-V'(q)j_x+\ell V'(q)j_\theta]
+\tau_{\rm a}J_\theta-fJ_x\nonumber\\
&=-\frac{\dd}{\dd t}\avg U+\tau_{\rm a}J_\theta-fJ_x.
\end{align}
At steady state $\dot S=0$ and $\dd\avg U/\dd t=0$, giving Eq.~\eqref{eq:power}.

\section{Stationary reduction and exact current formula}
\label{app:reduction}

The transformation $(x,\theta)\mapsto(q,\theta)$ has unit Jacobian.  The symmetry \eqref{eq:symmetry} and uniqueness of the stationary density imply Eq.~\eqref{eq:factorization}.  The noise in $q=x-\ell\theta$ is
\begin{equation}
\sqrt{2\mu_xT}\,\xi_x-\ell\sqrt{2\mu_\theta T}\,\xi_\theta
=\sqrt{2MT}\,\eta_q,
\end{equation}
with unit covariance because the original noises are independent.

To derive Eq.~\eqref{eq:Jq_exact}, write Eq.~\eqref{eq:Jq_local} as
\begin{equation}
\rho'(q)+\psi'(q)\rho(q)=-\frac{J_q}{MT}.
\end{equation}
Multiplication by $\e^{\psi(q)}$ gives
\begin{equation}
\frac{\dd}{\dd q}[\e^{\psi(q)}\rho(q)]
=-\frac{J_q}{MT}\e^{\psi(q)}.
\end{equation}
Integrating from $q$ to $q+L$, using $\rho(q+L)=\rho(q)$ and $\psi(q+L)=\psi(q)+\Lambda L/T$, yields Eq.~\eqref{eq:rho_exact}.  Integrating that density over one period gives Eq.~\eqref{eq:Jq_exact}.

The current relations follow from
\begin{align}
J_x&=\mu_x[-\avg{V'}-f],\qquad
J_\theta=\mu_\theta[\tau_{\rm a}+\ell\avg{V'}],\nonumber\\
v_q&=J_x-\ell J_\theta.
\end{align}
Solving these three linear equations for $J_x$ and $J_\theta$ gives Eqs.~\eqref{eq:Jx_vq} and \eqref{eq:Jth_vq}.

\section{Weak-coupling expansion}
\label{app:weak}

Let $V(q)=V_0\cos(kq)$ and expand
\begin{equation}
\rho(q)=\frac{1}{L}+V_0\rho_1(q)+O(V_0^2),
\qquad
J_q=-\frac{M\Lambda}{L}+O(V_0^2).
\end{equation}
The first-order part of Eq.~\eqref{eq:Jq_local} is
\begin{equation}
T\rho_1'(q)+\Lambda\rho_1(q)=\frac{k}{L}\sin(kq).
\end{equation}
The periodic, zero-mean solution is
\begin{equation}
\rho_1(q)=
\frac{k\Lambda}{L[\Lambda^2+(Tk)^2]}\sin(kq)
-\frac{Tk^2}{L[\Lambda^2+(Tk)^2]}\cos(kq).
\end{equation}
Since $v_q=-M[\Lambda+\avg{V'}]$ and
\begin{equation}
\avg{V'}=-kV_0^2\int_0^L\sin(kq)\rho_1(q)\,\dd q
=-\frac{k^2\Lambda V_0^2}{2[\Lambda^2+(Tk)^2]},
\end{equation}
Eq.~\eqref{eq:vq_weak} follows.  Because the free-coupling stall load is zero, $f_{\st}=O(V_0^2)$, so $\Lambda$ in the $O(V_0^2)$ correction may be evaluated at $f=0$, giving Eq.~\eqref{eq:fst_weak}.

\section{Step-by-step derivation of the stall reconstruction formula}
\label{app:reconstruction}

For completeness we collect, in one place, the steps that lead from the local reciprocal-current identity to the exact reconstruction formula~\eqref{eq:exact_reconstruction}, so the logic can be followed without cross-referencing the main text.  The physical content is simple: at mechanical stall the observed coordinate has zero mean motion, but its fluctuation--response violation still encodes the hidden-rotor dissipation.

\paragraph*{Local velocities and entropy production.}
In terms of the local mean velocities of Eq.~\eqref{eq:local_vel_def},
\begin{equation}
\nu_x(q)=\frac{j_x(q)}{p(q)},
\qquad
\nu_\theta(q)=\frac{j_\theta(q)}{p(q)},
\end{equation}
the steady entropy production~\eqref{eq:ep} is
\begin{equation}
\ep=\int_0^L\rho(q)
\left[\frac{\nu_x(q)^2}{\mu_xT}+\frac{\nu_\theta(q)^2}{\mu_\theta T}\right]\dd q.
\label{eq:app_ep}
\end{equation}
At mechanical stall the observed mean velocity vanishes, $J_x=\avg{\nu_x}=0$, so the visible dissipation is simply
\begin{equation}
\Phi_x^{\st}=\frac{\avg{\nu_x^2}}{\mu_xT}.
\label{eq:app_Phi}
\end{equation}

\paragraph*{Reciprocal local identity.}
The reciprocal force--torque structure gives the pointwise identity~\eqref{eq:stall_local_identity}, which at stall reads
\begin{equation}
\nu_\theta(q)=J_\theta^{\st}-\frac{\ell\mu_\theta}{\mu_x}\nu_x(q).
\label{eq:app_identity}
\end{equation}
The hidden local velocity therefore has two parts: a uniform hidden rotation $J_\theta^{\st}$ and a reaction part tied to the local motion in $x$.  Squaring,
\begin{equation}
\nu_\theta(q)^2
=(J_\theta^{\st})^2
-2J_\theta^{\st}\frac{\ell\mu_\theta}{\mu_x}\nu_x(q)
+\frac{\ell^2\mu_\theta^2}{\mu_x^2}\nu_x(q)^2,
\end{equation}
and averaging over one period with $\rho$; since stall gives $\avg{\nu_x}=0$, the cross term vanishes:
\begin{equation}
\avg{\nu_\theta^2}=(J_\theta^{\st})^2
+\frac{\ell^2\mu_\theta^2}{\mu_x^2}\avg{\nu_x^2}.
\label{eq:app_nuth_square}
\end{equation}

\paragraph*{Entropy decomposition at stall.}
Substituting Eq.~\eqref{eq:app_nuth_square} into Eq.~\eqref{eq:app_ep} at stall,
\begin{align}
\ep^{\st}
&=\frac{\avg{\nu_x^2}}{\mu_xT}
+\frac{1}{\mu_\theta T}
\left[(J_\theta^{\st})^2+\frac{\ell^2\mu_\theta^2}{\mu_x^2}\avg{\nu_x^2}\right]
\nonumber\\
&=\left(1+\frac{\ell^2\mu_\theta}{\mu_x}\right)
\frac{\avg{\nu_x^2}}{\mu_xT}
+\frac{(J_\theta^{\st})^2}{\mu_\theta T}.
\end{align}
With the reciprocal mobility factor $K=1+\ell^2\mu_\theta/\mu_x$ of Eq.~\eqref{eq:K} and the visible dissipation~\eqref{eq:app_Phi}, this is the decomposition~\eqref{eq:stall_decomp},
\begin{equation}
\ep^{\st}=K\Phi_x^{\st}+\frac{(J_\theta^{\st})^2}{\mu_\theta T},
\label{eq:app_decomp}
\end{equation}
which still contains the hidden current $J_\theta^{\st}$.

\paragraph*{Eliminating the hidden current.}
At stall the mechanical current vanishes, so the steady power balance~\eqref{eq:power} reduces to $T\ep^{\st}=\tau_{\rm a}J_\theta^{\st}$, while Eq.~\eqref{eq:hidden_stall_current} gives $J_\theta^{\st}=\mu_\theta(\tau_{\rm a}-\ell f_{\st})$, i.e.\ $\tau_{\rm a}=\ell f_{\st}+J_\theta^{\st}/\mu_\theta$.  Hence
\begin{equation}
T\ep^{\st}
=\left(\ell f_{\st}+\frac{J_\theta^{\st}}{\mu_\theta}\right)J_\theta^{\st}
=\ell f_{\st}J_\theta^{\st}+\frac{(J_\theta^{\st})^2}{\mu_\theta}.
\label{eq:app_power}
\end{equation}
Multiplying Eq.~\eqref{eq:app_decomp} by $T$ gives a second expression, $T\ep^{\st}=TK\Phi_x^{\st}+(J_\theta^{\st})^2/\mu_\theta$.  Comparing with Eq.~\eqref{eq:app_power} cancels the common quadratic term and yields the inference identity~\eqref{eq:inference_identity},
\begin{equation}
TK\Phi_x^{\st}=\ell f_{\st}J_\theta^{\st}
\quad\Longrightarrow\quad
J_\theta^{\st}=\frac{TK\Phi_x^{\st}}{\ell f_{\st}}
\qquad(\ell f_{\st}\neq0),
\label{eq:app_inference}
\end{equation}
in agreement with Eq.~\eqref{eq:Jth_reconstructed}.

\paragraph*{Back-substitution.}
Inserting Eq.~\eqref{eq:app_inference} into the decomposition~\eqref{eq:app_decomp},
\begin{equation}
\ep^{\st}=K\Phi_x^{\st}
+\frac{1}{\mu_\theta T}\left(\frac{TK\Phi_x^{\st}}{\ell f_{\st}}\right)^2
=K\Phi_x^{\st}
+\frac{TK^2(\Phi_x^{\st})^2}{\mu_\theta\ell^2f_{\st}^2},
\end{equation}
and using $\ell^2\mu_\theta=\mu_x(K-1)$ in the denominator gives the reconstruction formula~\eqref{eq:exact_reconstruction},
\begin{equation}
\ep^{\st}=K\Phi_x^{\st}
+\frac{TK^2(\Phi_x^{\st})^2}{\mu_x(K-1)f_{\st}^2}.
\end{equation}
The only inputs are the measured visible dissipation $\Phi_x^{\st}$, the stall load $f_{\st}$, and the bare constants $T$, $\mu_x$, $\ell$, $K$.

\end{document}